# Consistency checks of results from a Monte Carlo code intercomparison for emitted electron spectra and energy deposition around a single gold nanoparticle irradiated by X-rays

H. Rabus[1,12], W.B. Li[2,12], H. Nettelbeck[1,12], J. Schuemann[4,12], C. Villagrasa[3,12], M. Beuve[5,12], S. Di Maria[6,12], B. Heide[7,12], A.P. Klapproth[2,8], F. Poignant[5,9], R. Qiu[10,12], B. Rudek[4,11]

[1] *Physikalisch-Technische Bundesanstalt, Braunschweig and Berlin, Germany*
[2] *Institute of Radiation Medicine, Helmholtz Zentrum München - German Research Center for Environmental Health, Neuherberg, Germany*
[3] *Institut de Radioprotection et de Sûreté Nucléaire, Fontenay-Aux-Roses, France*
[4] *Massachusetts General Hospital & Harvard Medical School, Department of Radiation Oncology, Boston, MA, USA*
[5] *Institut de Physique des 2 Infinis, Université Claude Bernard Lyon 1, Villeurbanne, France*
[6] *Centro de Ciências e Tecnologias Nucleares, Instituto Superior Técnico, Universidade de Lisboa, Bobadela LRS, Portugal*
[7] *Karlsruhe Institute of Technology, Karlsruhe, Germany*
[8] *TranslaTUM, Klinikum rechts der Isar, Technische Universität München, Munich, Germany*
[9] *Present address: National Institute of Aerospace, Hampton, VA, USA*
[10] *Department of Engineering Physics, Tsinghua University, Beijing, China*
[11] *Present address: Perlmutter Cancer Center, NYU Langone Health, New York City, NY, USA*
[12] *European Radiation Dosimetry Group (EURADOS) e.V, Neuherberg, Germany*

**Abstract**

Organized by the European Radiation Dosimetry Group (EURADOS), a Monte Carlo code intercomparison exercise was conducted where participants simulated the emitted electron spectra and energy deposition around a single gold nanoparticle (GNP) irradiated by X-rays. In the exercise, the participants scored energy imparted in concentric spherical shells around a spherical volume filled with gold or water as well as the spectral distribution of electrons leaving the GNP. Initially, only the ratio of energy deposition with and without GNP was to be reported. During the evaluation of the exercise, however, the data for energy deposition in the presence and absence of the GNP were also requested. A GNP size of 50 nm and 100 nm diameter was considered as well as two different X-ray spectra (50 kVp and 100kVp). This introduced a redundancy that can be used to cross-validate the internal consistency of the simulation results. In this work, evaluation of the reported results is presented in terms of integral quantities that can be benchmarked against values obtained from physical properties of the radiation spectra and materials involved. The impact of different interaction cross-section datasets and their implementation in the different Monte Carlo codes is also discussed.

## 1. Introduction

Gold nanoparticles (GNPs) have been shown to enhance the biological effectiveness of ionizing radiation *in-vitro* and *in-vivo* (Hainfeld et al., 2004; Her et al., 2017; Cui et al., 2017; Kuncic and Lacombe, 2018; Bromma et al., 2020). This effect is often attributed to a dose enhancement due to the higher absorption of radiation by the high-Z material gold as compared to other elemental components of tissue. For example, the ratio of the mass-energy absorption coefficients of gold and soft tissue is between 10 and 150 for photons in the energy range between 5 keV and 200 keV (Butterworth et al., 2012). Due to Auger cascades following the creation of inner shell holes, a larger number of low-energy secondary electrons may lead to additional energy deposition in the vicinity of a GNP (McMahon et al., 2011). This results in an additional local enhancement of absorbed dose around a GNP, compared to the case when the GNP volume is filled with water. Since this local dose enhancement is limited to microscopic dimensions, Monte Carlo (MC) simulations are needed to determine its value.

Prompted by the large variety of results reported in literature regarding this dose enhancement (Mesbahi, 2010; Vlastou et al., 2020; Moradi et al., 2021), a code intercomparison exercise was organized as a joint activity of the Working Groups 6 "Computational Dosimetry" (Rabus et al., 2021a) and 7 "Internal Dosimetry" (Breustedt et al., 2018) of the European Radiation Dosimetry Group (Rühm et al., 2018, 2020). The exercise was an intercomparison of Monte Carlo simulations for the electron spectra emitted and the dose enhancement around a single GNP in water subject to X-ray irradiation. Two sizes (50 nm and 100 nm diameter) of spherical GNPs were irradiated by two different X-ray spectra (50 kVp and 100 kVp, for details see (Li et al., 2020a).

To emphasize the impact of differences between codes with respect to electron transport simulation and associated electron interaction cross sections, an artificial simple irradiation geometry was used: A parallel beam of photons emitted perpendicularly from a circular source area in the direction of the GNP. The diameter of the source was 10 nm larger than the GNP diameter, and it was located at 100 µm distance from the GNP center.

Participants in the exercise were to implement this geometry and the given photon energy spectra into their simulation and then report the following results for each combination of GNP size and X-ray spectrum: (a) the spectral distribution of electrons emitted from the GNP per primary photon emitted from the source, (b) the dose enhancement ratio (DER) in spherical shells around the GNP, i.e. the ratio of the energy deposited per primary photon in the presence and absence of the GNP. At a later stage of the exercise evaluation, participants were asked to report the energy deposition per primary photon for the simulations with and without the GNP.

The spherical shells used for scoring energy deposition had a thickness (difference between outer and inner radius) of 10 nm up to an outer radius equal to $r_g + 1$ µm, where $r_g$ is the GNP radius. Beyond this distance, 1 µm increments were used up to an outer radius of $r_g + 50$ µm.

First results from the exercise have been reported by Li et al. (2020a, 2020b) and the relation of the DER values with those relevant for realistic irradiation scenarios with extended photon beams have been discussed by Rabus et al. (2019, 2021b). This work focusses on the methodology used in the assessment of the reported results for consistency between the different cases (GNP sizes, X-ray spectra) and for consistency with the principle of energy conservation. These consistency checks allowed cases of improper implementation of the exercise to be detected. The influence of electron transport in the various MC codes is also discussed.

## 2. Materials and Methods

### 2.1. Criterion for consistency between integrals of the emitted electron spectra and deposited energy

The results from the two subtasks of the exercise, i.e. energy deposited around and emitted electron spectra from the GNP are complementary, as the extra energy deposited in the presence of the GNP is mainly imparted by interactions of electrons emitted from the GNP. For a quantitative comparison, this extra energy deposition around the GNP can be approximated by the difference between the energies imparted in the presence and absence of the GNP.

The first plausibility check was whether the difference of the reported energy deposition with and without the presence of the GNP (in spherical shells around the GNP) was compatible with the energy spectra of electrons emitted from the GNP.

To test this, one needs to consider (a) the total additional energy $\Delta E_{g,w}$ deposited in the presence of the GNP in the total scoring volume (i.e. a spherical shell of inner radius $r_g$ and outer radius $r_g+50$µm) per photon interaction and (b) the total energy $E_e$ transported out of the GNP by electrons. These two quantities were calculated from eqs. (1) and (2).

$$\Delta E_{g,w} = \sum_i \left[ \bar{\varepsilon}_g(r_i) - \bar{\varepsilon}_w(r_i) \right] \qquad (1)$$

where $\bar{\varepsilon}_g(r_i)$ and $\bar{\varepsilon}_w(r_i)$ are the average imparted energies (Booz et al., 1983) per primary photon in the $i$-th radial shell (with outer radius $r_i$) obtained in the simulations with and without the GNP, respectively.

$$E_e = \sum_j T_j \times N_E^{(e)}(T_j) \times \Delta T_j \qquad (2)$$

In eq. (2), $T_j$ and $\Delta T_j$ are the center and the width of the $j$-th energy bin of the electron spectra. $N_E^{(e)}$ is the distribution of particle number with respect to energy (Seltzer et al., 2011) of electrons leaving the GNP (i.e. number of electrons per energy interval, hereafter called spectral frequency).

From energy conservation, if all deposited energy is scored (i.e. for infinitely large outer radius of the scoring region), then $\Delta E_{g,w}$ should be almost the same as $E_e$. The ratio $\Delta E_{g,w} / E_e$ should be slightly smaller than unity since the spectrum of emitted electrons also includes those produced outside the GNP that subsequently traverse it. Furthermore, emitted electrons can be backscattered into the GNP where they subsequently deposit part of their energy.

## 2.2. Criteria for consistency between the data for different GNP sizes and photon energy spectra

The criteria outlined in the preceding section can be used to check the consistency between the electron spectra and energy deposition results for each combination of GNP size and photon spectrum. Consistency between results for different combinations of GNP size and photon spectrum can subsequently be achieved by using a different normalization of the results.

In the exercise, normalization was requested per primary photon. However, only a small fraction of the primary photons interacts in the GNP. The emitted electrons and extra energy deposition scored in the simulations is mainly due to cases where a photon interaction in the GNP occurs.

The expected number $\bar{n}_g$ of photon interactions in the GNP is approximately given by eq. (3).

$$\bar{n}_g = \frac{4\pi}{3} r_g^3 \int \mu_g(E) \Phi^{(p)}(E) e^{-\mu_w(E) d_s} dE \quad (3)$$

and depends on the GNP size and photon energy spectrum.

In eq. (3), $\mu_g(E)$ and $\mu_w(E)$ are the total linear attenuation coefficients of gold and water (Berger et al., 2010), respectively. $E$ is the photon energy, $r_g$ is the GNP radius, and $d_s$ is the distance of the GNP center from the photon source.

$\Phi^{(p)}(E)$ is the spectral fluence (particles per area and energy interval) of primary photons emitted from the source, which fulfills the normalization condition

$$\int \Phi^{(p)}(E) dE = \frac{1}{r_b^2 \pi} \quad (4)$$

where $r_b$ is the radius of the circular photon source used in the simulations. The values of $\bar{n}_g$ for the primary fluences used in the exercise are shown in Table 1.

Normalizing the quantities $\Delta E_{g,w}$ and $E_e$ by $\bar{n}_g$

$$\Delta E_{d,g}^* = \frac{\Delta E_{d,g}}{\bar{n}_g} \qquad E_e^* = \frac{E_e}{\bar{n}_g} \quad (5)$$

approximately gives the total energy $\Delta E_{g,w}^*$ deposited around a GNP in which a photon interaction occurred, and the total energy $E_e^*$ transported out of such a GNP by electrons.

The resulting second plausibility check was to test whether these two quantities were compatible with the average energy $E_{tr,g}$ transferred to electrons when a photon interacts with a gold atom. $E_{tr,g}$ depends on the photon energy spectrum and was calculated according to eq. (6).

**Table 1** Mean number of photon interactions in a GNP ($\bar{n}_g$) for the two GNP diameters and X-ray radiation qualities used in the exercise. The values apply to the fluences used for normalization of the results in the exercise (Li et al., 2020a). (1 photon per area of the photon source, i.e. per $2.8 \times 10^3$ nm² and $9.5 \times 10^3$ nm² for the 50 nm and 100 nm-diameter GNPs, respectively.)

|            | 50 kVp               | 100 kVp              |
|------------|----------------------|----------------------|
| 50 nm GNP  | $1.1 \times 10^{-3}$ | $5.4 \times 10^{-4}$ |
| 100 nm GNP | $2.6 \times 10^{-3}$ | $1.3 \times 10^{-3}$ |

$$E_{tr,g} = \frac{\int E \mu_{tr,g}(E) \Phi^{(p)}(E) e^{-\mu_w(E) d_s} dE}{\int \mu_g(E) \Phi^{(p)}(E) e^{-\mu_w(E) d_s} dE} \quad (6)$$

In eq. (6), $E$ is the photon energy, $\mu_{tr,g}$ is the energy transfer coefficient of gold, $\Phi^{(p)}$ is the particle fluence of primary photons emitted from the X-ray source, , and $d_s$ is the distance of the GNP center from the photon source. For evaluation of $E_{tr,g}$, $\mu_{tr,g}$ was approximated by the energy absorption coefficient $\mu_{en,g}$ taken from Hubbell and Seltzer (2016). (Strictly speaking, eq. (6) therefore gives a lower bound to the energy transferred to electrons, as they will lose some of their energy by bremsstrahlung collisions.)

As the electrons released in photon interactions with gold atoms lose part of their energy within the GNP before leaving it, the ratios $\Delta E_{g,w}^*/E_{tr,g}$ and $E_e^*/E_{tr,g}$ must be less than unity. Furthermore, the ratio should be smaller for the 100 nm GNP than for the 50 nm GNP (for the same photon spectrum), as the average path travelled by electrons before leaving the GNP is less for the smaller GNP.

For the same GNP size, the ratio for the 100 kVp spectrum should be smaller than for the 50 kVp spectrum, since the electrons produced by photo-absorption in the L, M, and outer shells as well as by Compton scattering have higher energies. The 100 kVp photon spectrum also contains photon energies where K shell absorption is possible. The fraction of such photons is, however, small and the photo-absorption coefficient around the K shell of gold is lower than in the photon energy range below 50 keV, where the majority of photons in the spectrum appear (Berger et al., 2010).

*2.3. Criterion for correct normalization*

A third plausibility check was based on the ratio of the total energy $E_{dep,w}(R)$ deposited in a water sphere of radius $R$ in the absence of GNPs to the average energy $E_{tr,w}(R)$ transferred by photon interactions in water (in the section of the sphere traversed by the primary photon beam). The latter is given by

$$E_{tr,w}(R) = D_w^{(p)} \rho_w r_b^2 \pi \times 2R \quad (7)$$

where the volume traversed by the beam is approximated by a cylindrical volume, $\rho_w$ is the density of water, $r_b$ is the radius of the photon beam, and $D_w^{(p)}$ is the average collision kerma. Owing to the small attenuation of the photon beam over the microscopic dimensions of the geometry, the mean collision kerma can be approximated by its value at the location of the GNP, which is calculated with eq. (8) using a primary photon spectral fluence $\Phi^{(p)}$ that satisfies eq. (4).

$$D_w^{(p)} = \int E \times \frac{\mu_{en,w}(E)}{\rho_w} \times \Phi^{(p)}(E) e^{-\mu_w d_s} dE \quad (8)$$

In eq. (8), $E$ is the photon energy, $\mu_{en,w}(E)/\rho_w$ is the mass energy absorption coefficient of water, $\Phi^{(p)}(E)$ is the spectral fluence of primary photons emitted from the source, $\mu_w(E)$ is the total linear attenuation coefficient of water and $d_s$ is the distance of the GNP's center from the photon source.

The deposited energy $E_{dep,w}(R)$ for $R=r_j$, where $r_j$ is the outer radius of the $j$-th spherical shell in the simulations, is approximately given by

$$E_{dep,w}(R = r_j) = \sum_{i=1}^{j} \bar{\varepsilon}_w(r_i) \quad (9)$$

With increasing $R$, the condition of longitudinal secondary electron equilibrium (i.e. along the direction of the primary photon beam) will be fulfilled, such that the ratio $E_{dep,w}(R)/E_{tr,w}(R)$ should converge with increasing $R$ to a value close to unity. The asymptotic value will not be unity as the simulation results also include energy deposited by electrons produced in interactions of photons that have been previously scattered out of the photon beam as well as any descendant photons. This effect leads to the value of $E_{dep,w}(R)$ being larger than $E_{tr,w}(R)$.

As the volume corresponding to the GNP was not used for scoring in the simulations, the value obtained by eq. (9) slightly underestimates the true value of $E_{d,w}(R)$. However, as this volume is less that $10^{-9}$ of the total volume, this can be considered negligible. Similarly, the fact that a sphere is used for scoring rather than a plane parallel slab will also lead to a slight reduction of $E_{dep,w}$ that should depend on the value of $R$. In fact, the deviation of the ratio $E_{dep,w}(R)/E_{tr,w}(R)$ from the saturation value followed an approximate $1/R$ dependence for $R \geq 30$ μm, such that the saturation value could be determined by linear regression of the ratio as a function of $1/R$.

*2.4. Final results of the exercise*

For the sets of results where the consistency tests indicated specific normalization issues, the respective participants were requested to check and confirm whether their simulations were compromised by the respective problem. Examples include improper implementation of the simulation geometry, such as using a source where the radius was larger than the GNP radius by 10 nm rather than the source diameter being 10 nm larger than the GNP diameter. If the participant confirmed that the simulations were biased as suggested by the outcomes of the consistency checks, the results were corrected accordingly.

As the energy binning of the electron spectra was not specified in the exercise definition, participants reported the spectra in different linear binning with bin widths ranging between 5 eV and 100 eV. Two participants used logarithmic binning with 100 intervals per decade. Consequently, the comparison of the spectra as reported by the participants in Fig. 7 of (Li et al., 2020a, 2020b) was compromised by the statistical fluctuations of the spectra reported with narrow energy bins.

All electron spectra reported with linear binning were therefore resampled such that a bin size of 100 eV was used up electron energies of 10 keV and a bin size of 500 eV beyond. As all linear bin widths were factors of 100 eV, a grouping of adjacent bins was possible. In addition, the distribution with respect to energy of the radiant energy (Seltzer et al., 2011) transported by the electrons (hereafter called spectral radiant energy) was also determined by calculating the ratio of the integral kinetic energy within each of the new kinetic energy bins to the width of the energy bin. The electron spectra reported in logarithmic binning have not been changed. The spectral radiant energy was determined in this case by multiplying the frequency per bin width by the arithmetic mean of the bin boundaries.

## 2.5. Participant identification and codes used

In this article, the participants of the exercise are identified by a letter (first letter in the name of the code used) and a number (if several participants used codes starting with the same letter). The rationale is that the discrepancies found in the evaluation of the exercise results cannot be attributed to the codes used but rather originate in most cases from incorrect implementation of the exercise definition in the simulations. To facilitate comparison with the reports of the preliminary results of the exercise in Li at al. (2020a, 2020b), a brief summary of the meaning of these labels is given below.

Participants G1, G2, and G3 all used GEANT4 with its low energy extensions and the track structure capabilities of GEANT4-DNA (Incerti et al., 2010; Bernal et al., 2015; Incerti et al., 2018) for simulating particle transport in water. Participants G1 and G3 used version 10.4.2, participant G2 version 10.0.5. The respective labels used in Li at al. (2020a, 2020b) were G4/DNA#1, G4/DNA#2, and G4/DNA#3.

Participant M1 used the 2013 release of MCNP6 (Goorley et al., 2012) version 6.1, participant M2 used MDM (Gervais et al., 2006), participant N used NASIC (Li et al., 2015) version 2018 and participant P1 used PARTRAC (Friedland et al., 2011) version 2015. In the work from Li at al. (2020a, 2020b), these participants were identified by the respective code names.

Participants P2 and P3, who both used PENELOPE (Salvat et al., 2011; Salvat, 2015), were identified as PENELOPE#1 and PENELOPE#2. Participant P2 originally used version 2011 for the simulations, while updated results were produced with the 2018 release. Participant P3, on the other hand, used the 2014 release of PENELOPE. Participant T, who used TOPAS-nBio version 1.0-beta with TOPAS version 3.1p3 (Schuemann et al., 2019), was identified as TOPAS.

## 3. Results and Discussion

### 3.1. Integrals of radial energy deposition around a GNP and energy spectra of ejected electrons

Fig. 1 shows a summary of all results reported by participants that have been evaluated in terms of the ratio $E_e^* / E_{tr,g}$ (ratio of the average energy transported by electrons leaving a GNP per photon interaction in the GNP to the mean energy released by a photon interaction in gold). The corresponding outcome of the evaluation in terms of $\Delta E_{g,w}^* / E_{tr,g}$ (ratio of the excess energy imparted around a GNP in which a photon interacts to the mean energy released by a photon interaction in gold) is shown in Fig. 2.

Preliminary results are indicated by superscripts on the participant identifier and have been withdrawn (&,#) or replaced by data obtained by correcting the normalization to the requested primary photon fluence (of one photon per source area). Participant G2 withdrew the electron spectrum results for the 100 nm GNP irradiated by the 50 kVp photon spectrum (for lack of explanation in failing the consistency checks) and provided new simulation results for the case of a 50 nm GNP and 50 kVp spectrum.

Participants P2 and P3 withdrew their results after realizing that in their simulations, the cumulative distribution had been mistakenly used for the probability distribution of the photon spectrum. Participant P2 repeated the simulations with the correct photon spectrum and, thus, provided revised solutions (Li et al., 2020b). Owing to limitations of the code used, the simulations had to be performed for a square-shaped photon source, but the respective fluence correction was applied to obtain the final results shown in Fig. 1 (and also in Fig. 2).

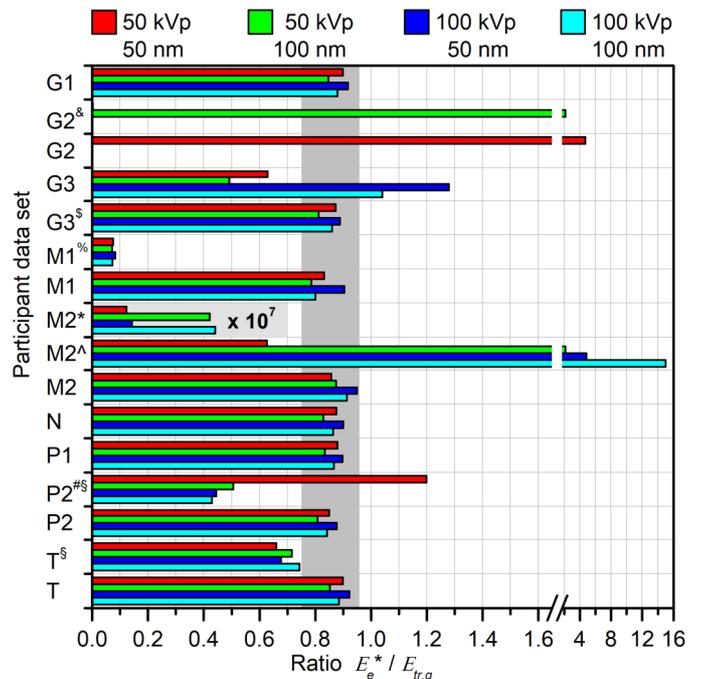

**Fig. 1:** Ratio of the total energy transported by electrons leaving a GNP that experienced a photon interaction to the mean energy transferred to electrons when a photon interacts in gold. The superscripts next to the participant identifiers indicate results where deviations from the exercise definition were revealed by the consistency checks and have been confirmed: § variation in simulation geometry (final results have been corrected); # variation in photon energy spectrum (results withdrawn); * variation in the normalization to primary particle fluence (final results have been corrected). The other superscripts indicate results that: % were obtained by using an incorrect tally for the angular range (and could be approximately corrected using a constant scaling factor); ^ were multiplied with incorrect factors to correct for particle fluence; & failed the consistency checks for unknown reasons and have been withdrawn; $ have been tentatively corrected for a suspected variation in simulation geometry (not confirmed by the participant).

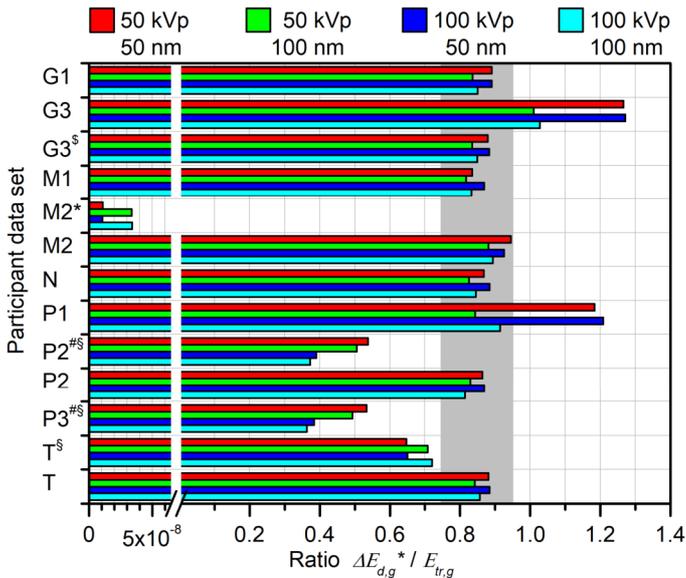

**Fig. 2:** Ratio of the total excess energy deposited around a GNP undergoing a photon interaction to the mean energy transfer to electrons when a photon interacts in gold. The grey shaded area indicates the expected range for this ratio. (See Fig. 1 for the meaning of the superscripts.)

The ensembles of results shown in Fig. 1 and Fig. 2 are different for several reasons: First, participant G2 only submitted results for electron spectra but not for energy deposition, while participant P3 only reported energy deposition but not electron spectra. Second, participant M1 used the wrong tally for scoring electrons leaving the GNP, but the correct one for scoring energy deposition so that these latter data were not updated. Third, at the time of the first report on the exercise (Li et al., 2020a) the bias of the results of participant M2 was only noticed for the electron spectra, since only the ratio of energy deposition with and without the GNP was requested. As the integral energy deposition in the absence of the GNP is insensitive to the chosen beam diameter (as long as it is small compared to the cross-section of the scoring volume), fewer results are shown in Fig. 3 compared to Fig. 1 and Fig. 2.

For all participants, the final results are those without superscript. With the exception of participants G2 and G3, these final results are all within the range expected from the principle of energy conservation that requires the values shown in Fig. 1 to be slightly smaller than unity, as a part of the energy transferred to electrons is absorbed in the GNP when a photon interacts there. This energy loss should be larger in the larger GNP and smaller for the higher-energetic X-ray spectrum. This expected behavior is observed for all results that fall in the expected range (indicated by the grey shaded area) with the exception of the results for participant M2. The reason for this exception could not be identified. The expected range was estimated based on the results reported by Koger and Kirkby (2016) and the uncertainties of the photon interaction coefficients (Andreo et al., 2012).

For participant G3, whose results failed the consistency checks, tentative results (G3$^\$$) are shown in Fig. 1 and Fig. 2 that would be obtained if (a) the reported data originated from simulations with a photon beam of equal diameter as the GNP and (b) the electron spectra from the 50 kVp X-ray spectrum are multiplied by a factor of 2 (as suggested by a comparison of the data for G3 in Fig. 1 and Fig. 2.)

As can be seen in both figures, these hypothetical corrections would make the results of participant G3 congruent with those of the other participants. However, as the participant could not confirm the suspected problems with the simulations, the reasons for the deviations remain unclear.

For the results of participant M2 in Fig. 1, a deviation of almost eight orders of magnitude from the results of other participants had been noticed in an early stage of the exercise and a potential reason and ensuing correction was suggested by the participant.

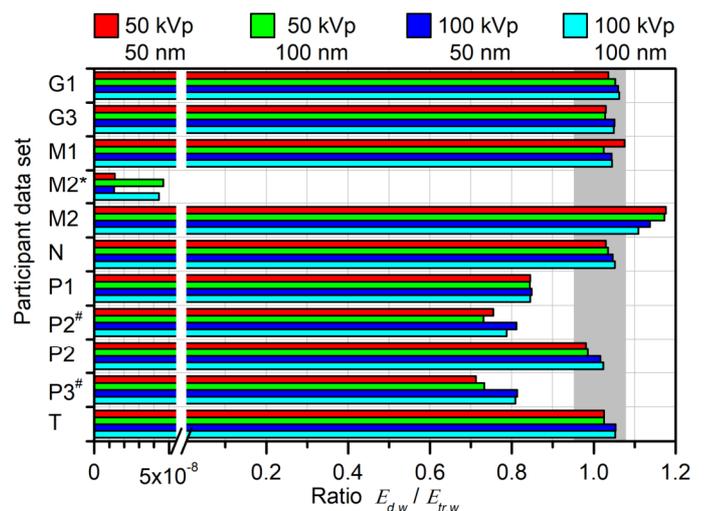

**Fig. 3:** Ratio of the energy deposited in the absence of the GNP summed over all spherical shells to the total energy transferred to electrons. This is for the case when a photon interacts in water within the section of the largest sphere that is traversed by the primary photon beam. A hashtag sign indicates data sets that were withdrawn by the participants.





Participant M2 did not simulate photon transport, but rather sampled from a uniformly distributed electron source (of energy distribution corresponding to the photon spectrum). The proposed correction was intended to correct the number of primary photons considered in the simulations. The data labelled as M2^ corresponds to the application of this proposed correction, which does not represent the data for M2 presented in (Li et al., 2020a) as such a correction was not correctly applied at that stage.

This bias of eight orders of magnitude also existed in the original results of M2 for energy deposition (see Fig. 2 and Fig. 3), but was not evident in the early stage of the exercise as only the DER was considered. The reason for this discrepancy was the use of a photon fluence of one particle per cm² instead of per source area (Li et al., 2020b). Additionally, the code used by participant M2 only scored energy deposition by ionizations and electronic excitations, which account for about 82% of the total imparted energy (Gervais et al., 2006). The data of participant M2 shown in Fig. 2 and Fig. 3 have been corrected accordingly.

The final data for M2 in Fig. 1 are based on electron spectra that deviate slightly from those shown in (Li et al., 2020b). This is due to inconsistencies in the data extraction from the results of participant M2 for the figures in (Li et al., 2020a). The results calculated from the correct data of participant M2 for emitted electrons, however, show a variation with photon spectrum and GNP size (Fig. 1 that disagrees with expected values (section 2.2): For the 50 kVp spectrum, the ratio $E_e^*$ / $E_{tr,g}$ increases with GNP size, where for both GNP sizes this ratio is smaller than $\Delta E_{g,w}^*$ / $E_{tr,g}$. Furthermore, the data of participant M2 shown in Fig. 3 are about 20% higher than the values that would be expected from the fact that this participant did not simulate photon transport. Since only electrons produced by photon interactions in the volume traversed by the primary photon beam were simulated, the data shown in Fig. 3 should be smaller than unity. This suggests further potential issues with the simulations of participant M2.

The results of the consistency checks also reveal a problem with the energy deposition results of participant P1: The values for energy deposition in the absence of the GNP are consistently a factor of about 0.8 too low (Fig. 3). This factor seems to be responsible for the systematic deviation of the DER values of participant P1 at large radial distances (50 µm) from the GNP shown in (Li et al., 2020a, 2020b). This deviation is approximately equal to the percentage of energy deposited in ionizations and electronic excitations.

However, this factor cannot be explained by such a partial scoring of deposited energy, since the ratio $\Delta E_{g,w}^*$ / $E_{tr,g}$ in Fig. 2 is about 1.2 for the 50 nm GNP and about 1 for the 100 nm GNP. The participant could not find an explanation for these observations.

For participants P2 and P3 a larger discrepancy can be seen in Fig. 3 as well as in Fig. 1 and Fig. 2. The origin of these discrepancies was the use of a different photon energy spectrum (Li et al., 2020b).

### 3.2. Internal consistency of simulation results

The energy transported by the electrons leaving the GNP and the additional energy deposited around it could also have been compared for each combination of photon spectrum and GNP size without prior normalization to the photon event frequency and without comparison with the expected energy transferred in a photon interaction.

This would have revealed inconsistencies between the simulations for energy deposition and for electron spectra such as observed for the 50 kVp results of participant G3.

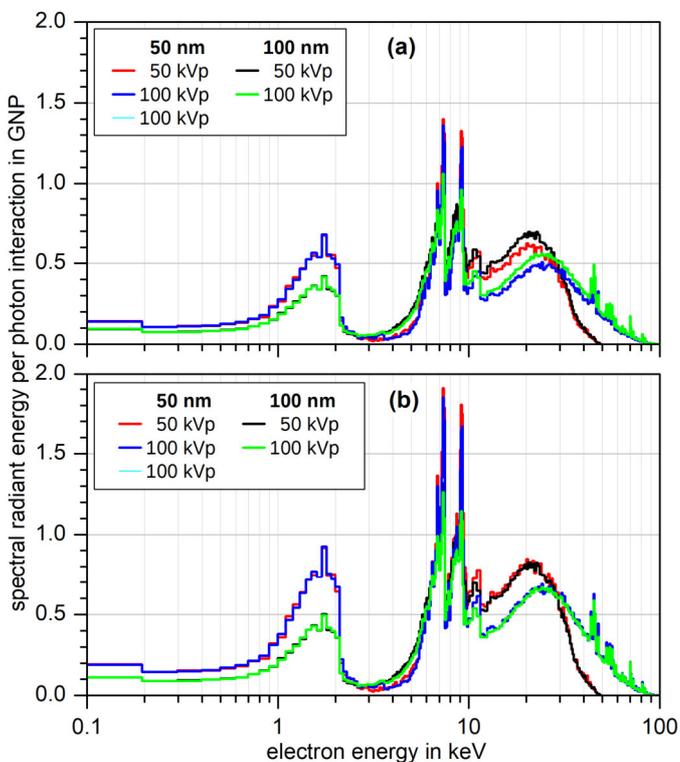

**Fig. 4:** Electron spectra reported by participant T for all combinations of GNP size and photon spectra (see legend). Data have been normalized to the number of photon interactions in the GNP expected for (a) beam diameter as defined in the exercise (GNP diameter plus 10 nm); (b) a beam radius equal to GNP radius plus 10 nm.





Detecting deviations from the defined geometry, however, requires at least a normalization to the GNP volume or the expected number of photon interactions in a GNP (eq. (3)). This is illustrated in Fig. 4 for the results of participant T, for which a comparison of Fig. 1 and Fig. 2 suggests consistency between the setups for electron spectra and energy deposition simulations. However, in both figures it can be seen that the data labelled by T$^§$ are significantly lower than the expected values (grey filled area). These data were obtained from the simulation results of participant T by normalizing to the expected number $\bar{n}_g$ of photon interactions (for the beam size of the exercise definition) and dividing by $E_{tr,g}$.

Fig. 4(a) shows the corresponding electron spectra of participant T rebinned and normalized to the expected number of photon interactions in the GNP for a photon fluence of one particle per circular source area (as per the exercise definition). In the supplementary Fig. S1, these data are compared with the originally reported finely binned results for the 50 nm GNP irradiated with the 100 kVp spectrum. It is evident from Fig. S1 that for energies above 10 keV the differences between the electron spectra for the same photon spectrum and different GNP size could not be detected with the narrow-binned spectra. As the energy loss due to interactions in the GNP is not significant for these high-energetic electrons, significant differences between the two GNP sizes are not plausible.

Fig. 4(b) shows the same data normalized to the expected number of photon interactions in the GNP for the source size used in the simulations of participant T. In this case, the expected agreement between data for the same photon spectra at high electron energies is observed. Furthermore, the difference between the spectra for different GNP sizes in the energy range of the M-shell Auger electrons (mostly between 1 keV and 2 keV) is also more pronounced. Here, the spectra for the different GNP sizes differ by roughly a factor of two as expected.

It should be noted that the quantity plotted on the *y*-axis in Fig. 4 is the spectral radiant energy transported

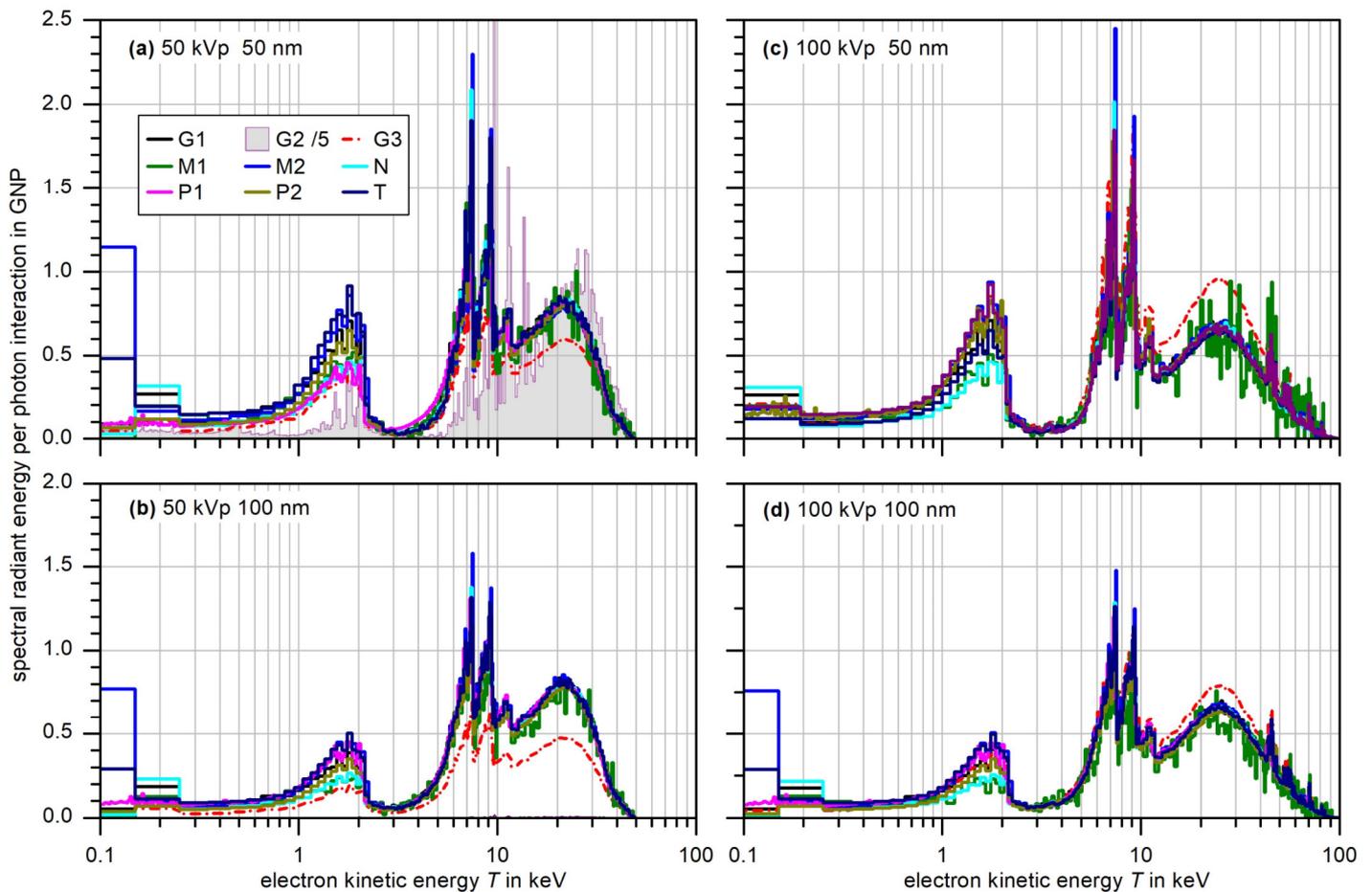

**Fig. 5:** Synopsis of the final spectral radiant energy of the electrons emitted from a GNP in which a photon interacts for the four cases studied in the exercise: (a) 50 kVp spectrum, 50 nm GNP, (b) 50 kVp spectrum, 100 nm GNP, (c) 100 kVp, 50 nm GNP; (d) 100 kVp spectrum, 100 nm GNP. The dot-dashed line and the shaded area represent datasets that failed the consistency checks. (Note that the data of participant G2 have been divided by a factor of 5.)





by the emitted electrons, i.e. the frequency in the respective energy bin multiplied by the energy of the bin center. As the *x*-axis is logarithmic, the area under the plotted curve represents the contribution of different energy ranges to the integral over all energies, i.e. the total number of electrons emitted from the GNP. In addition, the spectral shapes are more apparent than in Fig. 7 of (Li et al., 2020a, 2020b), where the details are hidden by the variation of frequencies over several orders of magnitude (and the fluctuations in the narrow-binned spectra).

The final results of all participants for the electron spectra are also presented in this way in Fig. 5. The data of participants G2 and G3 that failed the consistency checks have also been included. (The data of the former have been divided by a factor of 5 to fit the frame. For better visibility, they are shown here as shaded area rather than a dot-dashed line.) The results of all participants except these two are in good agreement at energies higher than 10 keV. For the regions of the Auger lines (below 2.2 keV and between 6 keV and 10 keV) significant differences are seen with the results deviating by factors of as much as two. The largest discrepancies can be seen in the energy range below 100 eV. Electrons in this energy range contribute negligibly to the total energy transported out of the GNP (see supplementary Fig. S2), but are relevant for the local dose increase in the proximity of the GNP (Rabus et al., 2021b).

### 3.3. Electrons ejected from a GNP

The presentation used in Fig. 5 highlights the spectral features of electron emission from a GNP. The variation in magnitude of the different participants results may reflect the impact of the different approaches used in the codes for simulating electron transport in gold and water. For a quantitative assessment of these differences, it is useful to consider the complementary integrals of the electron spectra:

$$\bar{n}_e^*(T_{min}) = \frac{1}{\bar{n}_g} \int_{T_{min}}^{T_{max}} N_E^{(e)}(T) dT \quad (10)$$

where $T$ is the kinetic energy of the electrons and $N_E^{(e)}$ is the number of emitted electrons per energy interval, $\bar{n}_g$ is the mean number of photon interactions in the GNP and $T_{max}$ is the highest possible electron energy. $\bar{n}_e^*(T_{min})$ is the average number of electrons emitted from a GNP experiencing a photon interaction that have a kinetic energy higher than $T_{min}$, which can be calculated directly from the electron spectra reported by the participants without the need for resampling. (This is also true for the total energy transported by

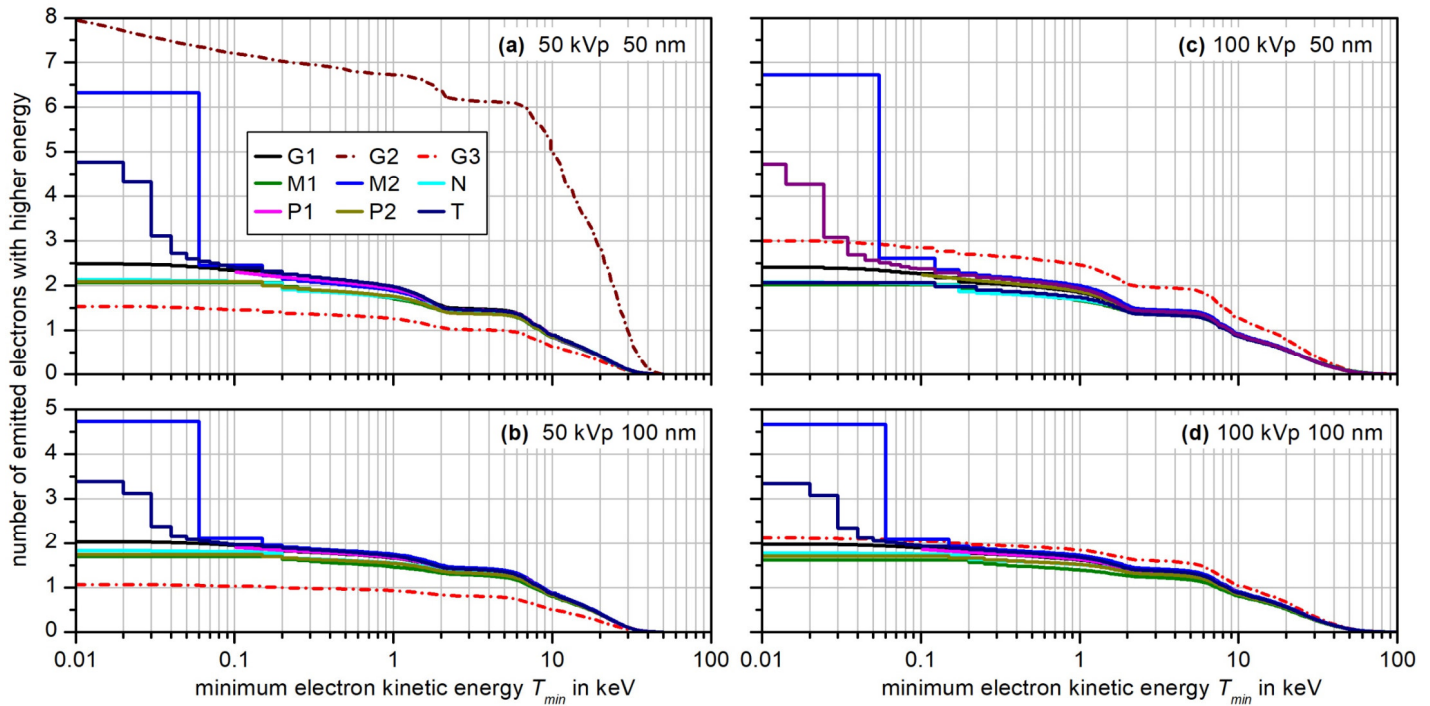

**Fig. 6:** Dependence of the number of electrons emitted from a GNP in which a photon interaction occurs that have a kinetic energy exceeding the value on the *x*-axis. (a) 50 kVp spectrum and 50 nm GNP, (b) 50 kVp spectrum and 100 nm GNP, (c) 100 kVp spectrum and 50 nm GNP; (d) 100 kVp spectrum and 100 nm GNP. Dot-dashed lines indicate data that failed the consistency checks. The different horizontal steps reflect the different bin sizes used by the participants.

electrons with kinetic energy exceeding $T_{min}$ as shown in Supplementary Fig. S3.)

The respective results are plotted in Fig. 6 such that the values are constant within an energy bin. Results that did not pass the consistency checks are shown as dot-dashed lines. It can be seen that for most spectra the predicted average number of electrons emitted after a photon interaction in a GNP is around 2. Only for participants M2 and T is this number significantly higher, where the discrepancy is primarily due to emitted electrons with energies below 100 eV. In the case of participant T this seems to be related to the use of a production threshold for secondary electrons as low as 10 eV. For participant M2, the increased number of low-energy electrons is presumably due to the fact that more than 1600 Auger and Coster-Kronig transitions were considered when simulating the de-excitation of ionized gold atoms. Furthermore, a newly developed electron cross-section dataset for gold (Poignant et al., 2020) was used in the code and the existence of a potential barrier at the GNP-water interface was also taken into account.

Comparison of Fig. 6(a) and (c) with Fig. 6(b) and (d), respectively, shows that the total number of electrons emitted is decreasing with increasing GNP diameter. Comparison of Fig. 6(a) and (b) with Fig. 6(c) and (d), respectively, reveals the number of emitted electrons is slightly smaller for the 100 kVp spectrum. Both observations are in agreement with the trends observed for the energy transported by leaving electrons.

A common observation in all four panels of Fig. 6 is that the results (apart from those of participants M2 and T) seem to fall into two groups that differ by about 10% with respect to the total number of emitted electrons. This is further illustrated in Fig. 7 where the integrals over energy ranges are shown for all combinations of GNP size and photon spectrum. The respective right-most histogram in each panel corresponds to the electron energy range above the highest Auger electron energy from an L-shell vacancy. With the exception of the results of participant G3 that failed the consistency checks, the values all scatter within 3% to 4% around an average value of about 0.75 for the 50 kVp spectrum and 0.8 for the 100 kVp spectrum. This seems reasonable given that only a fraction of the photons (with energies of 23 keV or higher) can produce L-shell photoelectrons of these energies, which is higher for the 100 kVp spectrum. Furthermore, there is also a significant probability for elastic photon scattering in the energy ranges considered in the exercise.

The histograms second from the right correspond to the energy range between 5 keV and 11.5 keV, where

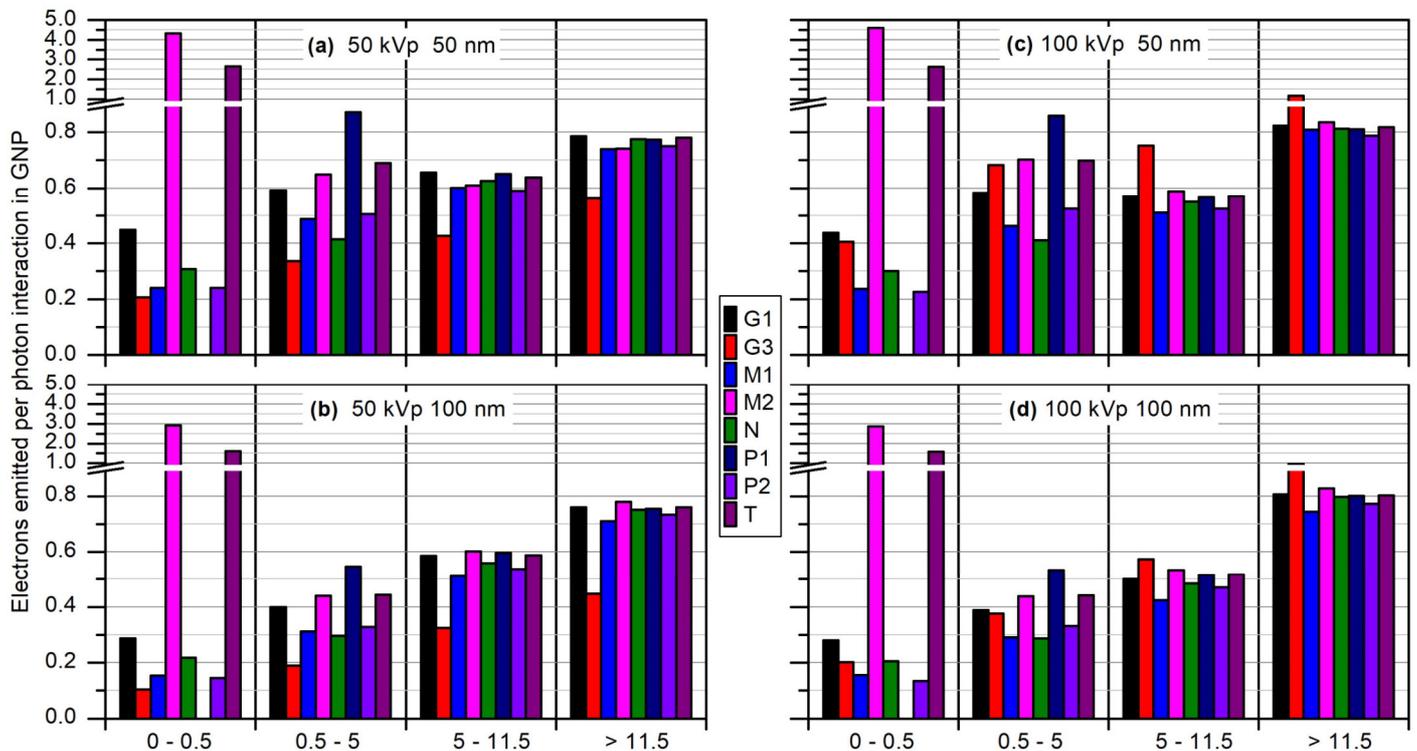

**Fig. 7:** Comparison of the integrals of the emitted electron spectra over different electron energy ranges (given on the abscissa in keV) for (a) 50 kVp spectrum and 50 nm GNP, (b) 50 kVp spectrum and 100 nm GNP, (c) 100 kVp and 50 nm GNP; (d) 100 kVp spectrum and 100 nm GNP.(The missing column in the left panel of each graph is due to the fact that participant P1 only reported electron energies higher than 100 eV.)

Auger-electrons are produced from L-shell vacancies filled by transitions involving only electrons from higher shells. In these histograms, the scatter is larger and the results show a dependence on GNP size and photon spectrum, that becomes evident when Fig. 7(a) and Fig. 7(d) are compared. These dependencies are more pronounced in the energy range between 500 eV and 5 keV, which covers Auger electrons from M-shell vacancies (and from L-shell vacancies filled with another L-shell electron). The scatter between results of different participants is most pronounced in the left-most histograms that cover the energy range below 500 eV.

Reference to the list of simulation parameters and cross-section datasets used by the participants in Table 1 of (Li et al., 2020a) does not provide a simple explanation for the differences observed in Fig. 6 and Fig. 7. The high number of low-energy electrons reported by participant T is most likely due to the low energy threshold for electron production. For participant M2, the high numbers may be due to comprehensive Auger cascades.

Nevertheless, the number of electrons emitted per photon interaction in the GNP that have energies greater than the highest L-shell or the highest M-shell Auger electron energy may also be used as a criteria for checking the consistency of simulated electron spectra from GNPs.

On the contrary, it is the low-energy region of the electron spectrum that is sensitive to simulation details such as interaction cross-sections, energy thresholds, and the scope of the transitions considered in relaxation processes following the creation of inner shell vacancies. The influence of procedures for particle transport, particularly across interfaces, is also greater in the low energy range. For instance, a surface potential barrier leads to a change of kinetic energy when the electrons cross the interface, and it also changes (reduces) their emission probability (Bug et al., 2012). This illustrates the need for a detailed investigation of these aspects in the frame of future intercomparison exercises.

It is worth noting in this context that most codes only consider atomic relaxation where the final state is a multiple charged ionized atom. In reality, all vacancies in valence shells of a GNP are filled and all holes are collected in the conduction band. The transitions leading to this final state also produce electrons with low energy (with respect to the Fermi edge) that may overcome the surface energy barrier.

## 4. Conclusion

The consistency tests presented in this paper have been used to identify simulation results that did not fully comply with the definition of the Monte Carlo code intercomparison exercise. Deviations from the exercise definition included variation in geometrical dimensions, different particle fluence, incorrect tallies and variations in the photon energy spectra. In the first two cases, the results could be corrected by a simple fluence correction. The other cases required determination of appropriate correction factors by performing additional simulations or repeating the simulations in the exercise. The cross-checking of internal consistency of the simulation results emphasizes the need for such multi-group intercomparison studies such as to raise awareness in the scientific community that apparent simplicity of a simulation task can be deceptive.

Apart from identifying inconsistencies between different simulations, the methods used in this study provide tools for assessing the plausibility of simulations results for the physical radiation effects of nanoparticles. Such plausibility checks are often not considered in such simulation studies reported in the literature (Rabus et al., 2021b).

In particular, normalizing the simulation results to the probability for a photon interaction in a GNP yields easily interpretable quantities. An example shown in this work was the total number of ejected electrons from a GNP. For the GNP sizes considered in the exercise, there are approximately two electrons with energies exceeding 100 eV that leave a GNP after a photon interaction. Electrons of lower energy will be absorbed in the few nm-thick coating of the GNPs. Thus, any radiation effects of GNPs of this size are due to only a few emitted electrons.

**Acknowledgements**

This work was, in part, funded by the DFG (grant nos. 336532926 and 386872118) and the National Cancer Institute (grant no. R01 CA187003). Werner Friedland is acknowledged for providing his simulation results without claiming co-authorship.

**Supplemental material to "Monte Carlo code intercomparison for emitted electron spectra and energy deposition around a single gold nanoparticle irradiated by X-rays" by H. Rabus *et al.*, Radiation Measurements, 2021**

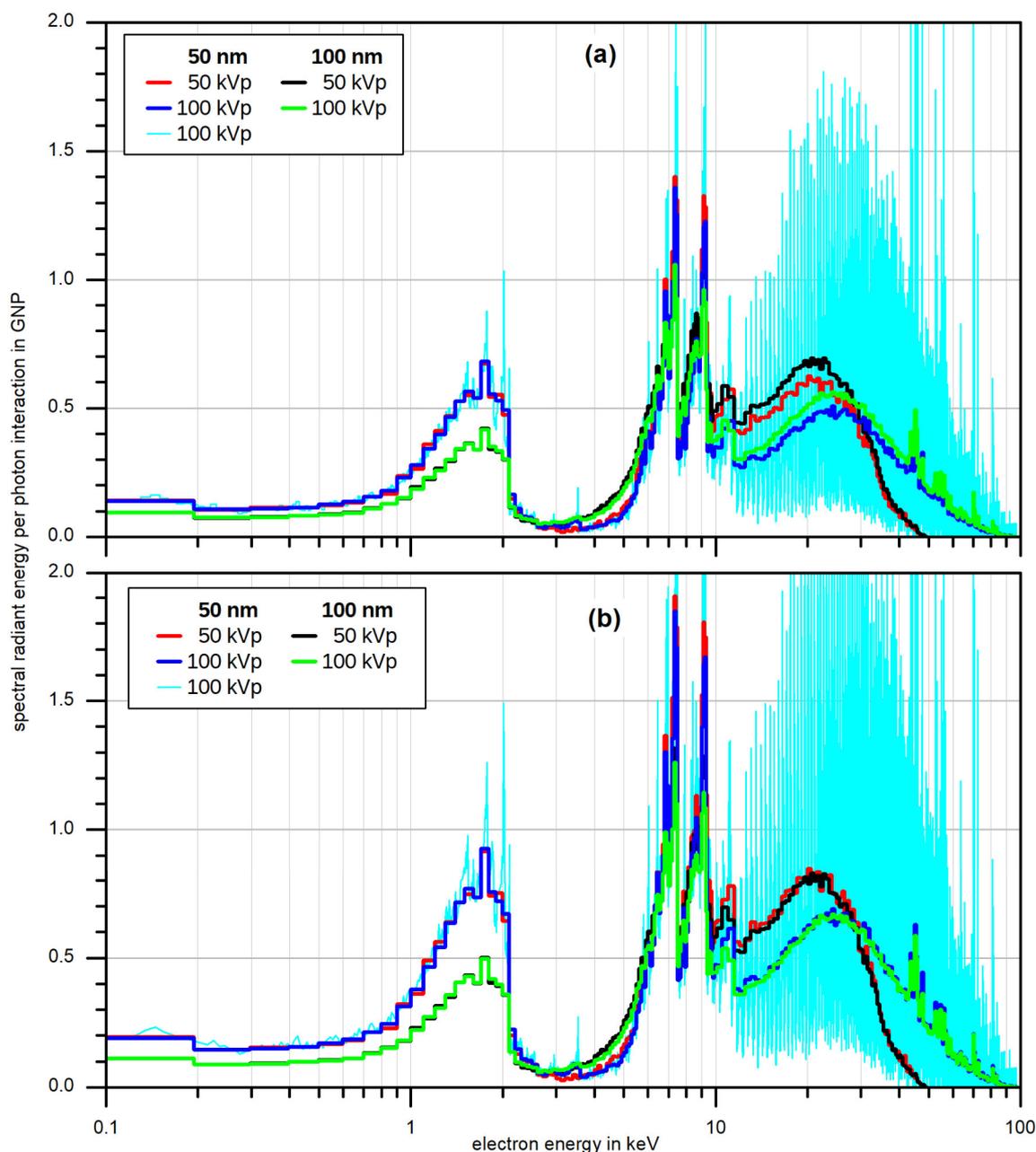

**Fig. S1:** Original electron spectra reported by participant T for the 50 nm GNP and 100 kVp photon spectrum (thin cyan lines) and rebinned spectra (as shown in **Fehler! Verweisquelle konnte nicht gefunden werden.**) for all combinations of GNP size and photon spectra (see legend). Data have been normalized to the number of photon interactions in the GNP expected for (a) beam diameter as defined in the exercise (GNP diameter plus 10 nm); (b) a beam radius equal to GNP radius plus 10 nm.

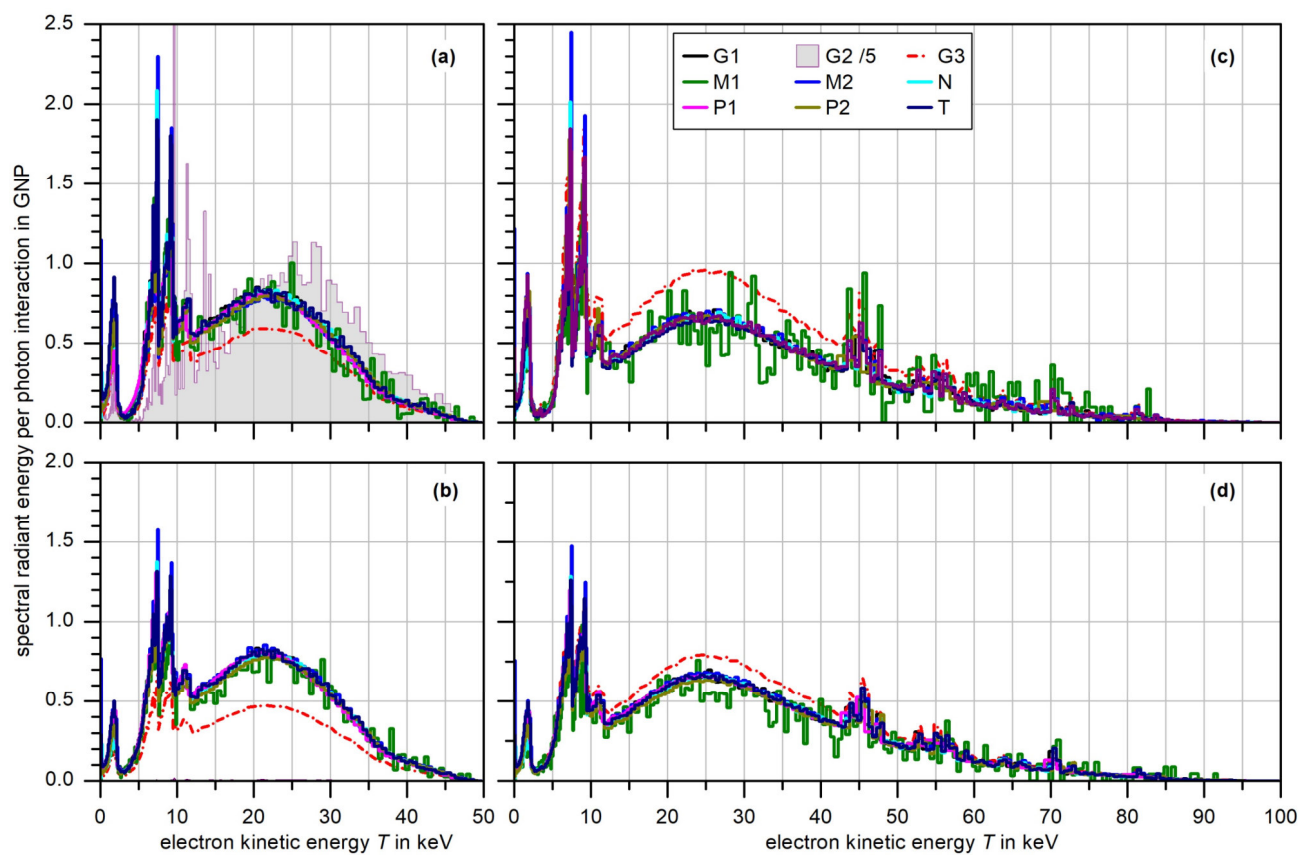

**Fig. S2:** Synopsis of the final radiant energy spectra of the electrons emitted from a GNP in which a photon interacts for the four cases studied in the exercise: (a) 50 kVp spectrum and 50 nm GNP, (b) 50 kVp spectrum and 100 nm GNP, (c) 100 kVp and 50 nm GNP; (d) 100 kVp spectrum and 100 nm GNP. The data of participant M1 have been multiplied by a correction factor of 11. Although the results of participants G2 (only for 50 kVp and 50 nm) and G3 failed the consistency checks, the data is included for completeness. The data of participant G2 have been divided by a factor of 5 and are shown as a shaded area rather than a dot-dashed line for better visibility. It should be noted that the *x*-axis is linear so that the area under the curves is proportional to the contribution of the respective energy range to the total energy transported by emitted electrons.

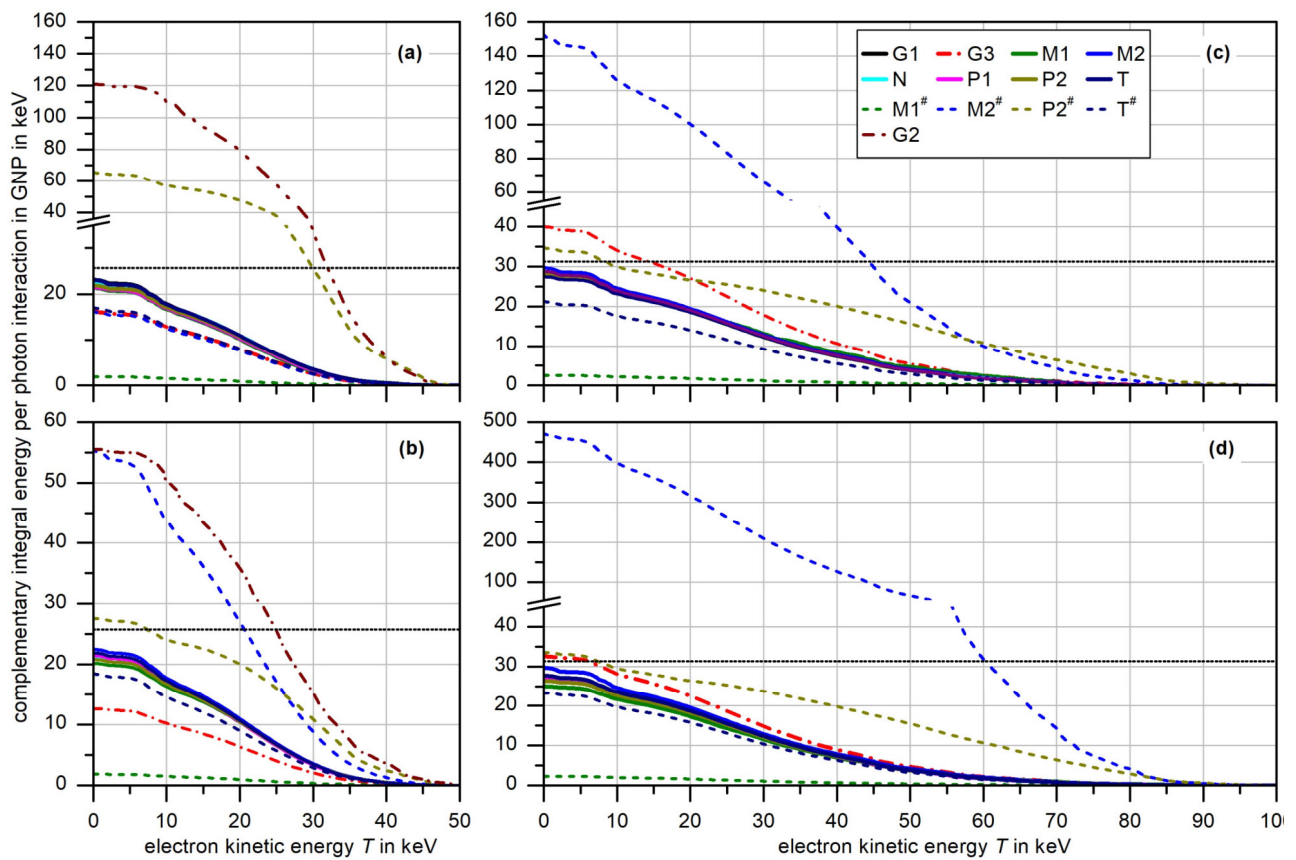

**Fig. S3:** Energy transported by electrons that are emitted from a GNP (in which a photon interacts) that have an energy exceeding the value plotted on the *x*-axis. The data have been derived from the reported data of participants (with their respective energy binning) for the four cases studied in the exercise: (a) 50 kVp spectrum, 50 nm GNP, (b) 50 kVp spectrum, 100 nm GNP, (c) 100 kVp, 50 nm GNP; (d) 100 kVp spectrum, 100 nm GNP. Dashed lines indicate data that have been superseded by new or corrected results. Dot-dashed lines indicate data failing the consistency checks that have not been revised.